\documentclass[journal=jacsat,manuscript=article]{achemso}

\usepackage[version=3]{mhchem} 

\usepackage{color}
\usepackage{graphicx}
\usepackage{caption}
\usepackage{subcaption}
\usepackage{comment}
\usepackage{gensymb}

\author{Giacomo Melani}
\affiliation{Department of Chemistry, Universit{\"a}t Z{\"u}rich, 8057 Z{\"u}rich, Switzerland}
\alsoaffiliation{Pritzker School of Molecular Engineering, University of Chicago, 60637, Chicago, US}
\altaffiliation{Equal contribution}
\email{giacomomelani@uchicago.edu}
\author{Juan Pablo Guerrero-Felipe}
\affiliation{Physics Department and IRIS Adlesrshof, Humboldt-Universität zu Berlin, 12489 Berlin, Germany}
\alsoaffiliation{Department of Physics, Free University Berlin, 14195 Berlin, Germany}
\altaffiliation{Equal contribution}
\author{Ana M. Valencia}
\affiliation{Institute of Physics, Carl-von-Ossietzy Universit{\"a}t Oldenburg, 26129 Oldenburg, Germany}
\author{Jannis Krumland}
\affiliation{Physics Department and IRIS Adlesrshof, Humboldt-Universität zu Berlin, 12489 Berlin, Germany}
\author{Caterina Cocchi}
\affiliation{Institute of Physics, Carl-von-Ossietzy Universit{\"a}t Oldenburg, 26129 Oldenburg, Germany}
\alsoaffiliation{Physics Department and IRIS Adlesrshof, Humboldt-Universität zu Berlin, 12489 Berlin, Germany}
\email{caterina.cocchi@uni-oldenburg.de}
\author{Marcella Iannuzzi}
\affiliation{Department of Chemistry, Universit{\"a}t Z{\"u}rich, 8057 Z{\"u}rich, Switzerland}
\email{marcella.iannuzzi@chem.uzh.ch}

\title{Donors, Acceptors, and a Bit of Aromatics: Electronic Interactions of Molecular Adsorbates on hBN and MoS$_2$ Monolayers}

\begin{document}

%
%

\newpage
\begin{abstract}
The design of low-dimensional organic-inorganic interfaces for the next generation of opto-electronic applications requires an in-depth understanding of the microscopic mechanisms ruling electronic interactions in these systems.
In this work, we present a first-principles study based on density-functional theory inspecting the structural, energetic, and electronic properties of five molecular donors and acceptors adsorbed on freestanding hexagonal boron nitride (hBN) and molybdenum disulfide (\ce{MoS2}) monolayers.
All considered heterostructures are stable, due to the crucial contribution of dispersion interactions, which are maximized by the overall flat arrangement of the physisorbed molecules on both substrates.
The level alignment of the hybrid systems depends on the characteristics of the constituents.
On hBN, both type-I and type-II heterostructures may form, depending on the relative energies of the frontier orbitals with respect to the vacuum level. 
On the other hand, all \ce{MoS2}-based hybrid systems exhibit a type-II level alignment, with the molecular frontier orbitals positioned across the energy gap of the semiconductor. 
The electronic structure of the hybrid materials is further determined by the formation of interfacial dipole moments and by the wave-function hybridization between the organic and inorganic constituents.
These results provide important indications for the design of novel low-dimensional hybrid materials with suitable characteristics for opto-electronics.

\end{abstract}

\newpage
\section{Introduction}

Hybrid materials formed by carbon-conjugated molecules adsorbed on low-dimensional semiconductors and insulators have been attracting attention due to the their structural versatility and electronic tunability.~\cite{Lee2014,zhen+16nano,breu+16pssrrl,Gobbi2018, dauk+19apx,mrky+19apl,rija+20jpcl,qiao+212DM,amst+21jpcl} Depending on their density on the substrate and on their physico-chemical characteristics, physisorbed moieties can introduce localized electronic states,~\cite{chou+17jpcc,zhon+18jpcl,wang-paul20pccp} dispersive bands,~\cite{rija+20jpcl} or a combination thereof.~\cite{cai+16cm,jing+jmca,krum-cocc21es} The electronic structure of the interface results from the level alignment between the organic and inorganic components~\cite{zhu+18sa,zhan+18am,aden-liu21jcp,park+21as,ye+21jpcl} and the hybridization between their electronic wave-functions.~\cite{song+17nano,shen-tao17ami,xie+19jpca,krum-cocc21es,guo+22nr} As both these effects depend on the intrinsic nature of the building blocks, the need for systematic analyses on the electronic structure of hybrid systems are in high demand. 

Electronic structure calculations based on density-functional theory (DFT) are particularly suited for this purpose~\cite{quek-khoo14acr,hofm+21pccp} and for exploring various material combinations without requiring empirical parameters. With the electron density being its central quantity, DFT grants immediate access to the charge redistribution induced by adsorption.~\cite{cai+16cm,jing+jmca,song+17nano,park+21am}
This way, it is possible to assess the type of ground-state doping and to gain insight into the spatial extension of the electron cloud at the interface. Furthermore, DFT calculations are able to deliver work functions, level alignments, band structures, and (projected) density of states, among other important properties.~\cite{cai+16cm,jing+jmca,zhen+16nano,park+21am,krum-cocc21es}
While state-of-the-art first-principles methods to obtain the electronic structure of solid-state materials are currently based on many-body perturbation theory,~\cite{drax+14acr,aden-liu21jcp} the choice of range-separated hybrid functionals to approximate the exchange-correlation potential in DFT offers the optimal trade-off between accuracy and computational costs.~\cite{park+21am,krum-cocc21es}
Proper inclusion of van der Waals interactions improves the prediction of structural arrangements and hence the description of electronic properties.~\cite{tkat+10mrs}
The level of accuracy currently achieved by such \textit{ab initio} calculations ensures reliable results complementary to experiments.~\cite{zhen+16nano,liu+17nl,park+21am}

In this work, we present a DFT study on the structural, energetic, and electronic properties of five representative organic molecules, including donor and acceptor compounds as well as a purely aromatic moiety, adsorbed on freestanding hexagonal boron nitride (hBN) and molybdenum disulfide (\ce{MoS2}) monolayers. 
The former is a known insulator, widely used as a substrate and/or as an encapsulating material in low-dimensional heterostructures,~\cite{zhan+18am} which has been receiving increasing attention in surface and interface science~\cite{Auwarter2012, Lin2012, Gomez2013, Weng2016, Zhang2017, Kim2018, Auwarter2019}, for instance to sustain the growth of well-defined organic thin films.~\cite{krat+19jpd,matk+19afm,amst+21jpcl} \ce{MoS2} belongs to the family of transition-metal dichalcogenides, the most promising emerging class of low-dimensional semiconductors.
By performing geometry optimizations using the generalized-gradient approximation (GGA) and refining the analysis of the electronic structure using a range-separated hybrid functional, we rationalize how the nature of the constituents of the hybrid interface determines the level alignment and the projected density of states. Our findings offer useful indications to interpret and predict the electronic properties of similar low-dimensional hybrid interfaces from the character of substrates and adsorbates.

\section{Methods and Systems}

\subsection{Computational details}\label{Methods}

All results presented in this work are obtained from DFT~\cite{hohe-kohn64pr} electronic structure calculations through the solution of the Kohn-Sham equations.~\cite{kohn-sham65pr}
The structures are optimized at the GGA level of theory, using the Perdew-Burke-Ernzerhof (PBE) functional.~\cite{perd+96prl}
To compute electronic properties on each optimized structure, including densities of states, energy levels alignment and molecular orbitals, the Heyd–Scuseria–Ernzerhof (HSE06)~\cite{heyd+06} range-separated hybrid functional is adopted.

For all complexes with hBN as a substrate, we employ the Gaussian and plane-wave formalism, as implemented in  the CP2K package.~\cite{cp2k2020} We choose the short-range-double-$\zeta$ MOLOPT basis sets~\cite{molopt2007} for the expansion of the valence electron density, while the interaction with the atomic cores is represented by  Godecker-Teter-Hutter (GTH) pseudopotentials.~\cite{GTH1996, GTH1998, GTH2005} 
The expansion of the density in an auxiliary plane waves basis is truncated at the kinetic-energy cutoff of 600~Ry.
The van der Waals (vdW) contributions are included either according to the Grimme-D3 scheme~\cite{grim06JCC} or by augmenting the exchange-correlation functional with the self-consistent rVV10 functional,~\cite{rvv10} which  in combination with PBE is known to provide reliable structural properties for similar hybrid interfaces.~\cite{Iannuzzi2014}
We apply the quasi-Newtonian  Broyden–Fletcher–Goldfarb–Shanno (BFGS) algorithm to find the minimum on the potential energy surface, with a convergence criterion of 5$\times$10$^{-4}$~Ha/bohr in the energy gradients. The Brillouin zone is sampled at the $\Gamma$-point only.

For the \ce{MoS2}-based interfaces, we use the plane-wave expansion of the wave-functions and the electron density as implemented in the Quantum Espresso (QE) code,~\cite{gian+17jpcm} with a cutoff of 30 and 300 Ry respectively, and the projector augmented-wave method.~\cite{bloc94prb} BFGS optimization is carried out with a threshold for the interatomic forces of 5$\times$10$^{-4}$~Ha/bohr.
A uniform 6$\times$6$\times$1 \textbf{k}-point mesh is adopted to sample the Brillouin zone and vdW corrections are included according to the Grimme-D3 scheme.~\cite{grim06JCC}

\subsection{Model systems}
We consider two-dimensional (2D) hybrid interfaces formed by five carbon-conjugated organic molecules physisorbed on monolayer hBN and \ce{MoS2}.
The organic molecules considered in this study exhibit different electronic characteristics: tetrathiafulvalene (TTF) and 2,2'-bithiophene (2T) are known to act as donors, while 7,7,8,8-tetracyanoquinodimethane (TCNQ) and its tetrafluorinated derivative (\ce{F4-TCNQ}) are strong acceptors;~\cite{sun+19am} for comparison, we additionally consider pyrene,~\cite{picc+19jpcl,herp+21jpca} a polycyclic aromatic hydrocarbon~\cite{dias85acr} of similar size as the aforementioned molecules. 
The hybrid model interfaces are constructed by placing one molecule on top of the two dimensional material with parallel backbone with respect to the substrate and running a geometry optimization (see Fig.~\ref{fig_bn_1}
and \ref{fig_mos2_1}). 
hBN is modelled in a 6$\times$6 supercell, where we adopted the experimental lattice constant for the unit cell, $a=2.5$~\AA{}.  
For \ce{MoS2}, we used a 4$\times$4 supercell with unit-cell lattice parameter $a = 3.19$~{\AA}. 
A sufficiently large amount of vacuum (20~\AA{} with \ce{MoS2} and 40~\AA{} with hBN) above the interfaces prevents spurious interactions between the periodic replicas under the applied periodic boundary conditions.

\section{Results and Discussion}
\subsection{Structural Properties}\label{structures}

\begin{figure}[h!]
  \centering
  \includegraphics[width=0.9\textwidth]{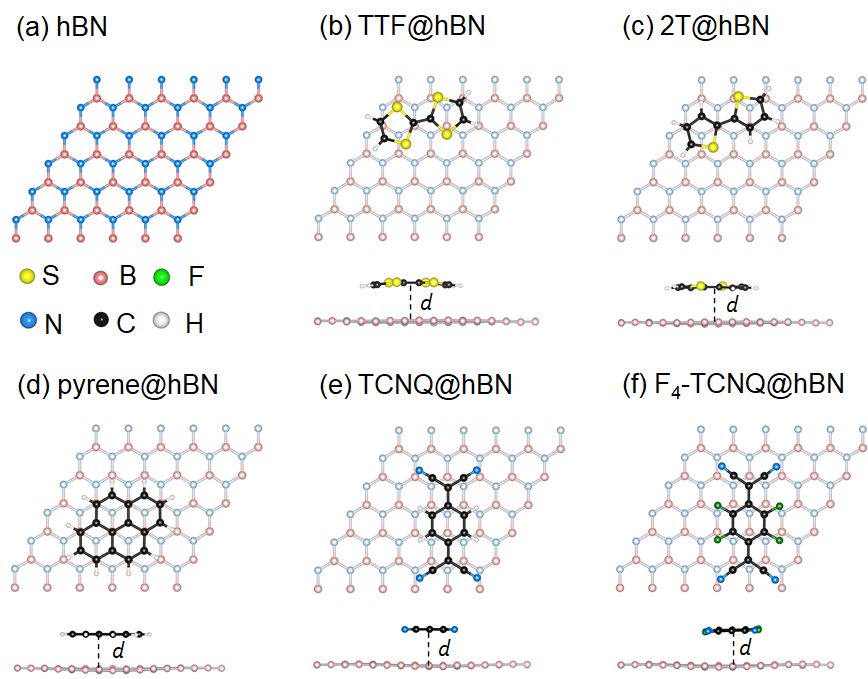}
\caption{ {\small (a) A 6$\times$6 supercell of hBN monolayer; top and side views of the hybrid interfaces formed by (b) tetrathiafulvalene (TTF), (c) bithiophene (2T), (d) pyrene, (e) tetracyanoquinodimethane (TCNQ), and (f) fluorinated TCNQ (\ce{F4-TCNQ}) adsorbates.}}
\label{fig_bn_1}
\end{figure}

All molecules adsorb approximately flat on top of hBN, thus maximising dispersion interactions.
The distance of the molecular species from the substrate plane ranges from 3.3 to 3.4~{\AA}.  
Upon adsorption, the molecular structures do not change appreciably compared to the gas-phase configurations. Exceptions concern a concave bending of TTF of about 11$^{\circ}$ towards the substrate (see Fig.~\ref{fig_bn_1}b), in contrast with previous results on metallic surfaces,~\cite{Wang2011, Kretz2021} where the molecule bends in a convex fashion due to the strong interactions with the metal electronic charge density. Furthermore, the 2T undergoes a backbone ``twist'' with a dihedral angle of 7$^{\circ}$ (Fig.~\ref{fig_bn_1}c).  
Finally, hBN is subject to a slight rippling as a result of the attractive $\pi$-$\pi$ interactions with the physisorbed molecules. 
Corresponding values of 0.21~\AA{}, 0.23~\AA{}, 0.27~\AA{}, and 0.29~\AA{} are found in the heterostructures with TTF and 2T, with TCNQ, with \ce{F4-TCNQ}, and with pyrene, respectively, see Fig.~\ref{fig_bn_1}.

\begin{figure}[h!]
  \centering
  \includegraphics[width=0.9\textwidth]{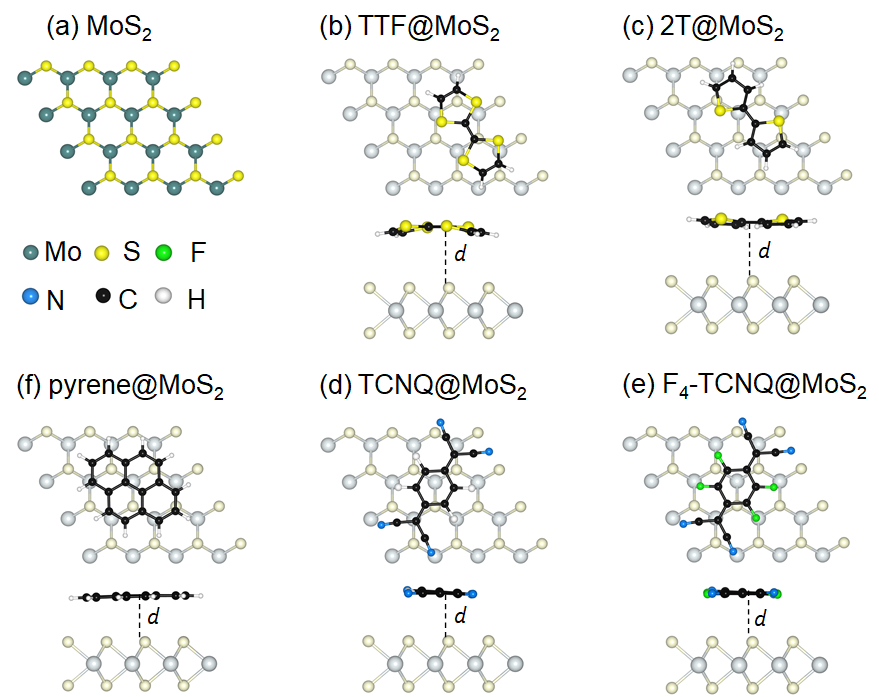}
\caption{ {\small (a) A 4$\times$4 supercell of \ce{MoS2} monolayer; top and side views of hybrid interfaces formed by (b) tetrathiafulvalene (TTF), (c) bithiophene (2T), (d) pyrene, (e) tetracyanoquinodimethane (TCNQ) and (f) fluorinated TCNQ (\ce{F4-TCNQ}) adsorbates.}}
\label{fig_mos2_1}
\end{figure}

In the hybrid heterostructure including the \ce{MoS2} substrate (Fig.~\ref{fig_mos2_1}a), the donor molecules TTF and 2T exhibit the same concave bending and backbone twisting as in the hBN-based ones discused above (see Fig.~\ref{fig_mos2_1}b,c). As a result, in these molecules, the hydrogen atoms are closer to the substrate than the carbon atoms, at a distance of 3.39~\AA{}.
The acceptors TCNQ and \ce{F4-TCNQ} are slightly bent, too, when physisorbed on \ce{MoS2}, with the nitrogen atoms pointing towards the substrate and being displaced 0.2~\AA{} downwards with respect to the backbone plane lying at 3.39~\AA{} above the monolayer (see Fig.~\ref{fig_mos2_1}e,f). This behavior is analogous to the one exhibited by these molecules on ZnO,~\cite{xu+13prl} on graphene,~\cite{kuma+17acsn} and on the hydrogenated Si(111) surface.~\cite{wang+19aem,jaco+20apx}
Finally, pyrene, which is planar in the gas phase,~\cite{dias85acr} remains such also upon adsorption, and lays at a distance of 3.32~\AA{} from \ce{MoS2}.

\subsection{Energetics}
\begin{table*}
  \caption{Adsorption energy ($E_\text{ads}$) calculated for the hBN-based heterostructure at the PBE-vdW level, using both the D3 and rVV10 scheme for the vdW contributions; interaction energy ($E_\text{int}$) and dispersion energy ($E_\text{disp}$) computed at the PBE-rVV10 level of theory.  All values are in eV.}
  \label{tbl:example}
  \begin{centering}
  \begin{tabular}{lcccc}
    \hline
    System & $E_{\text{ads}}$(D3) & $E_{\text{ads}}$(rVV10) & $E_{\text{int}}(rVV10)$ & $E_{\text{disp}}(rVV10)$\\
    \hline
    TTF@hBN & -0.96 & -1.06 & -1.08 & -1.09 \\
    2T@hBN & -0.85 & -0.96  & -0.98 & -1.02\\
    Pyrene@hBN & -1.13 & -1.30 & -1.33 & -1.44 \\
    TCNQ@hBN  & -1.04 & -1.21  & -1.24 & -1.31 \\
    \ce{F4-TCNQ}@hBN & -1.14 & -1.41 & -1.45 & -1.49 \\
    \hline
  \end{tabular}
  \label{tab_bn_ads}
\end{centering}
\end{table*}

In order to quantify the energetic stability of the considered hybrid heterostructures, we introduce the adsorption energy defined as:
\begin{equation}\label{eq_ads}
    E_{\text{ads}} = E^{\text{opt}}_{\text{mol@surf}} - E^{\text{opt}}_{\text{surf}} - E^{\text{opt}}_{\text{mol}}, 
\end{equation}
where the superscript ``opt'' refers to the optimized  geometries and the subscripts ``mol'' and ``surf'' stand for the molecular and surface total energies, respectively.
In the hBN-based interfaces, the adsorption strength increases from donor-like systems to the acceptors (see Table~\ref{tab_bn_ads}) with \ce{F4-TCNQ}, the most electron-withdrawing molecule among the considering ones, leading to the most stable heterostructure precisely on account of this characteristic.~\cite{Greber2018, Auwarter2019}
To better characterize  the nature of the molecule-substrate interactions in the considred hybrid interfaces, it is convenient to single out the dispersion contribution from the interaction strength, by introducing the interaction energy
\begin{equation}\label{eq_int}
    E_{\text{int}} = E^{\text{opt}}_{\text{mol@surf}} - E_{\text{surf}} - E_{\text{mol}},
\end{equation}
where $E_{\text{surf}}$ and $E_{\text{mol}}$ are the single-point energies computed for the individual subsystems taken with the same coordinates as in the optimized complex. The dispersion contribution to each term is defined as the energy difference at fixed coordinates between a calculation with the vdW correction and one without it. The final contribution to the adsorption is given by the dispersion energy, defined as:
\begin{equation}\label{eq_disp}
    E_{\text{disp}} = E^{\text{disp}}_{\text{mol@surf}} - E^{\text{disp}}_{\text{surf}} - E^{\text{disp}}_{\text{mol}}.
\end{equation}
As expected, in the case of the hBN-based heterostructures, the dispersion contribution turns out to be predominant (see Table~\ref{tab_bn_ads}), confirming that no chemical bond is formed between the molecules and the monolayer. The small, yet noticeable, differences between adsorption and interaction energies (20--40 meV) indicate that the charge distribution and also the original geometries of both molecules and substrate are slightly perturbed upon physisorption. 
When comparing the interaction energy with the dispersion contribution, one observes that the latter is slightly more negative. This result points to a minor destabilization effect due to distortions upon molecular adsorption.
Indeed, the interaction term must include some repulsive (Pauli) contributions owing to the overlap of the electronic distributions of molecule and substrate, whereas the dispersion part is purely attractive.    
 
 Depending on the choice of vdW functional, the relative magnitude of dispersion \textit{vs.} interaction may somewhat vary, but our comparison between two approaches demonstrates the same qualitative picture (see Table~\ref{tab_bn_ads}). 
In both cases, all adsorption energies lie between  -0.9 and -1.4~eV and the relative trends in stability are the same.
The following electronic-structure calculations involving the hBN substrate are then restricted to the rVV10 approach only, which proved to yield reliable adsorption and structural properties.~\cite{Iannuzzi2014}
The adsorption of  TCNQ and TTF on hBN was  investigated in a previous work by Tang and coworkers~\cite{Tang2011} who applied DFT with the PBE functional and no additional vdW correction. The resulting adsorption energies are -0.112~eV and -0.041~eV, respectively, \textit{i.e.}, significantly weaker due to the missing dispersion contribution.

\begin{table}
\begin{tabular}{  l  c c c }
\hline
 System & $E_{\text{ads}}$(D3) & $E_{\text{int}}(D3)$ & $E_{\text{disp}}(D3)$ \\ \hline
TTF@MoS$_2$ &  -0.91 & -0.91 & -0.94 \\
2T@MoS$_2$ &  -0.77  & -0.77 & -0.82\\
Pyrene@MoS$_2$ & -1.02 & -1.02 & -1.17  \\
TCNQ@MoS$_2$ &  -0.88 & -0.88 & -0.97 \\
\ce{F4-TCNQ}@MoS$_2$ &  -0.97 & -0.97 & -1.01\\
\hline
\end{tabular}
\caption{Adsorption energy ($E_\text{ads}$), interaction energy ($E_\text{int}$) and  dispersion energy ($E_\text{disp}$) for the \ce{MoS2}-based heterostrustructures computed at the PBE-Grimme-D3 level of theory. All values are in eV.}
 \label{table:abs_energy}
\end{table}

Moving now to the \ce{MoS2}-based interfaces, we find a qualitatively similar trend in the adsorption energies as the one discussed above for the heterostructures with hBN (see Table~\ref{table:abs_energy}).
Among the considered systems, the least stable one is 2T@\ce{MoS2}, due to the twisted backbone of the molecule that reduces the attractive $\pi$-$\pi$ interactions with the substrate. Unsurprisingly, the most negative value of $E_{\text{ads}}$ is found for pyrene, which adsorbs flat on \ce{MoS2} (see Fig.~\ref{fig_mos2_1}d).
On the other hand, in all \ce{MoS2}-based heterostructures, adsorption and interaction energies exhibit differences on the order of 10$^{-3}$~eV, as a sign of negligible energy relaxation of the molecules and of the \ce{MoS2} monolayer when the hybrid interfaces are formed. These variations are one order of magnitude smaller than those computed for the hBN-based interfaces (see Table~\ref{tab_bn_ads}).
A reason for these contrasting behaviors can be ascribed to the chemical nature of the two substrates: While hBN is characterized by a N-rich surface, \ce{MoS2} has instead a S-rich one. Such bare distinction in the composition of the two inorganic materials affects the affinity of the adsorbates towards them. Indeed, N-containing molecules such as TCNQ and its fluorinated sibling adsorb more favorably on hBN than TTF and 2T which are S-rich, likely as a consequence of orbital overlap between atoms of same kind. 
The values of dispersion energies shown in Table~\ref{table:abs_energy} also exhibit a qualitative difference with respect to their counterparts in Table~\ref{tab_bn_ads}, namely, the dispersion contribution for pyrene on \ce{MoS2} is larger than the one for \ce{F4-TCNQ}.
This behavior which can be explained again based on the chemical affinity argument presented above.


\subsection{Electronic Properties}\label{electronic}

\begin{figure}
    \centering
    \includegraphics[width=0.9\textwidth]{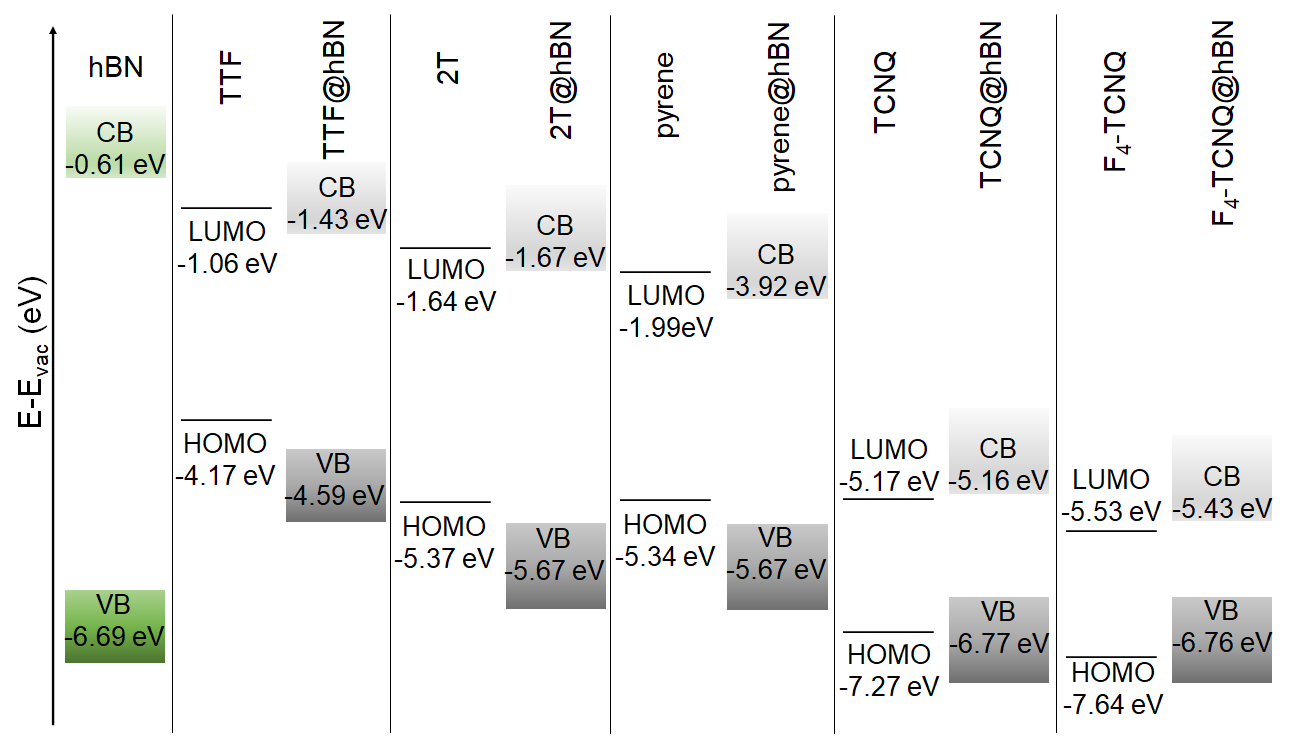}
    \caption{Energy level alignment computed for the hBN-based hybrid interfaces using the HSE06+rVV10 hybrid functional.}
\label{fig_bn_la}
\end{figure}

In the last part of our analysis, we inspect the electronic properties of the considered hybrid interfaces analyzing in particular the energy level alignment and the projected density of states. 
Again, we start from the hybrid systems including hBN. 
Like its bulk counterpart~\cite{Watanabe2004,blas+95prb,arna+06prl,aggo+18prb} monolayer hBN is an insulator~\cite{elia+19natcom} with a computed value of the quasi-particle band-gap above 7~eV.~\cite{galv+16prb,pale+182dm}
Our result obtained from DFT with the HSE06 hybrid functional (6.08~eV, see Fig.~\ref{fig_bn_la}) underestimates that value but it significantly improves it with respect to the one obtained from local DFT.~\cite{pale+182dm}
The agreement with experimental references is also very good.~\cite{Watanabe2004, Cassabois2016, Auwarter2019}
The large electronic gap of hBN and the absolute energies of its band edges determine the alignment with respect to the molecular frontier levels (Fig.~\ref{fig_bn_la}).
Both frontier states of TTF, 2T, and pyrene fall within the energy gap of hBN, leading to a type-I lineup.
In these three interfaces, the band edges lie within the band-gap of hBN, however, they are systematically downshifted by a few hundreds of meV with respect to the frontier states of the molecules.
In the interfaces including TCNQ and \ce{F4-TCNQ}, instead, the highest-occupied molecular orbital (HOMO) of the gas-phase molecules lies below the valence-band maximum (VBM) of free-standing hBN, giving rise to a type-II level alignment. 
In these cases, the highest occupied (lowest-unoccupied) level of the hybrid interface is downshifted (upshisfted) by a few tens of meV with respect to the respective counterpart in the isolated monolayer (molecule), see Fig.~\ref{fig_bn_la}.

\begin{figure}[h!]
\includegraphics[width=0.9\textwidth,clip=true]{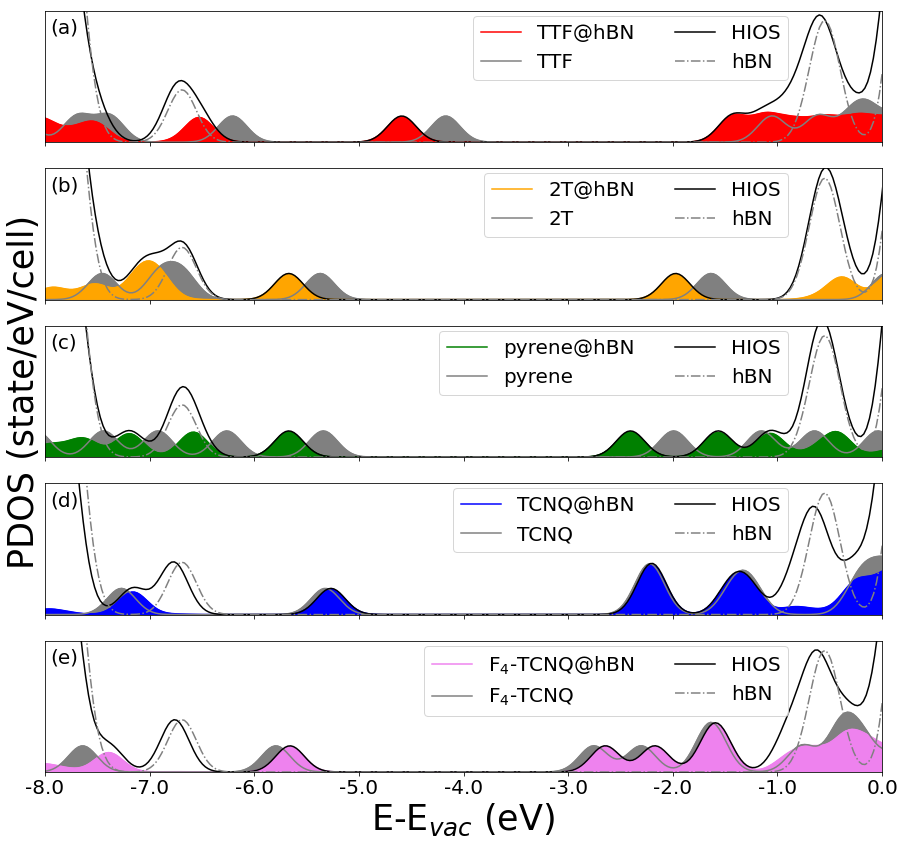}
\caption{Projected density of states for the hBN-based hybrid inorganic-organic systems (HIOS, black solid lines), including (a)-(b) the donors, TTF and 2T, (c) the aromatic molecule pyrene, and (d)-(e) the acceptors, TCNQ and \ce{F4-TCNQ}, calculated at the HSE06+rVV10 level of theory and compared against the results obtained for the isolated constituents shown by dashed lines (hBN) and gray areas (molecules).The contributions of the molecules within the hybrid interfaces are depicted by colored areas. A broadening of 500 meV is applied in all plots. The energy scale is offset to the vacuum level ($E_{vac}$).}
\label{fig_bn_pdos}
\end{figure}

The plots of the projected density of states (PDOS) reported in Fig.~\ref{fig_bn_pdos} confirm the picture rendered by Fig.~\ref{fig_bn_la}. Furthermore, they visually show that the localization of the frontier states reflects the energetic lineup of the electronic levels.
For a more detailed analysis, we include in Fig.~\ref{fig_bn_pdos} also the density of states of the isolated constituents.
For further comparison, the contributions of the molecules within the electronic structure of the hybrid interfaces are shown, too.
By inspecting these results, we identify two concomitant effects in the PDOS of the heterostructures.
First, the energy levels of the physisorbed molecules undergo a shift with respect to their counterparts in gas-phase. 
As the direction of this shift depends on the electron-donating (downwards) or -accepting (upwards) character of the molecule, we can rationalize this effect in terms of charge transfer.
With the moiety releasing or withdrawing electrons to or from the substrate, an interfacial dipole is formed.
For the chosen molecules, the electron-donating character of the donor is stronger in magnitude than the withdrawing ability of the acceptors.
As a result, the frontier levels of TTF, 2T, and pyrene are subject to a downshift of a few hundreds of meV, up to 0.5~eV; those of TCNQ and its perfluorinated counterpart undergo instead an upshift of the order of 100~meV.
The second effect disclosed from Fig.~\ref{fig_bn_pdos} is the electronic hybridization between the molecular orbitals and the hBN bands, which are particularly evident in the valence region of TTF@hBN, 2T@hBN, and, to a lesser extent, of pyrene@hBN (Fig.~\ref{fig_bn_pdos}a-c), as well as in the conduction region of the interfaces hosting the molecular acceptors (Fig.~\ref{fig_bn_pdos}d-e). 
With the partial exception of the lowest-unoccupied molecular orbital (LUMO) of TTF and the HOMO of \ce{F4-TCNQ}, the frontier states of the hybrid systems do not hybridize with the hBN bands. 

\begin{figure}[h!]
    \centering
    \includegraphics[width=0.9\textwidth]{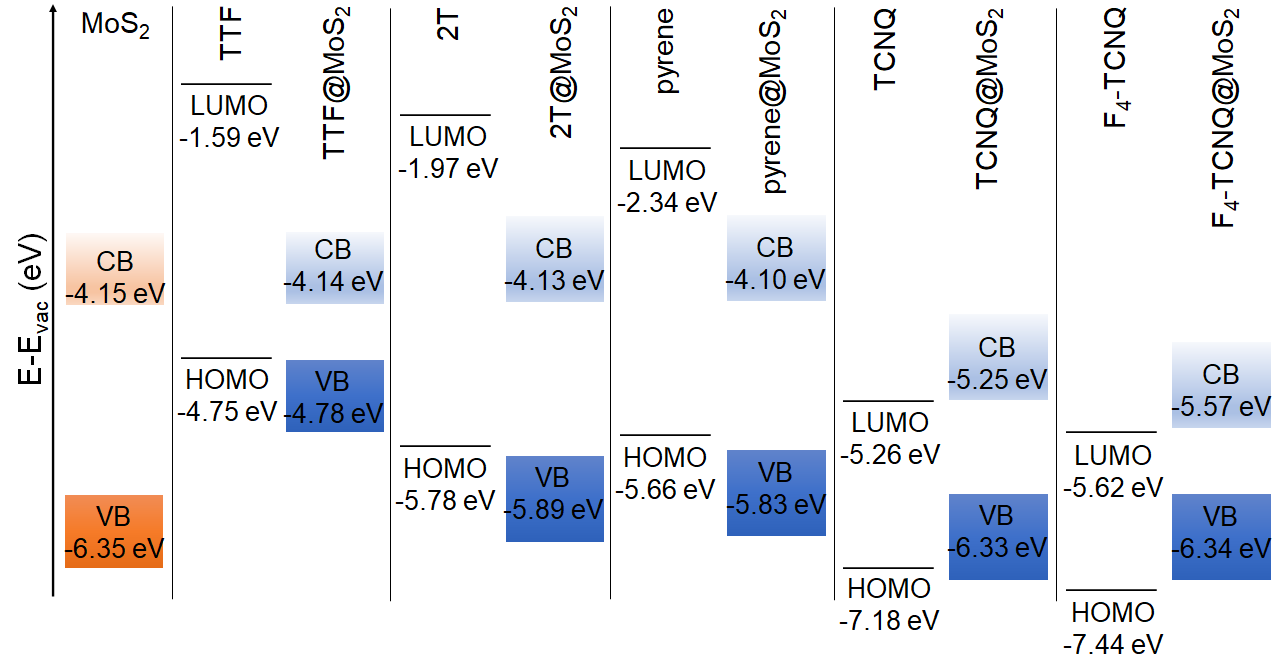}
    \caption{Energy level alignment computed for \ce{MoS2}-based hybrid interfaces using the HSE06+D3 functional.}
\label{Fig:LA-MoS2}
\end{figure}

Moving now to the electronic properties of the \ce{MoS2}-based hybrid interfaces, we notice that all these systems exhibit a type-II level alignment, with the band edges of the heterostructures being determined by the electron-donating ability of the absorbed molecule (see Fig.~\ref{Fig:LA-MoS2}).
Upon adsorption of TTF, 2T, and pyrene, the highest-occupied state of the interface corresponds to the HOMO of the adsorbate, whereas the lowest-unoccupied one is given by the conduction-band minimum (CBM) of free-standing \ce{MoS2}. 
On the contrary, the LUMO of the molecular acceptors, TCNQ and \ce{F4-TCNQ}, falls within the energy gap of \ce{MoS2}, whereas the HOMO of these molecules is lower than the VBM of the 2D material, in agreement with the known behavior of electron accepting molecules on this type of substrates.~\cite{jing+jmca}. 

\begin{figure}[h!]
    \centering
    \includegraphics[width=0.9\textwidth]{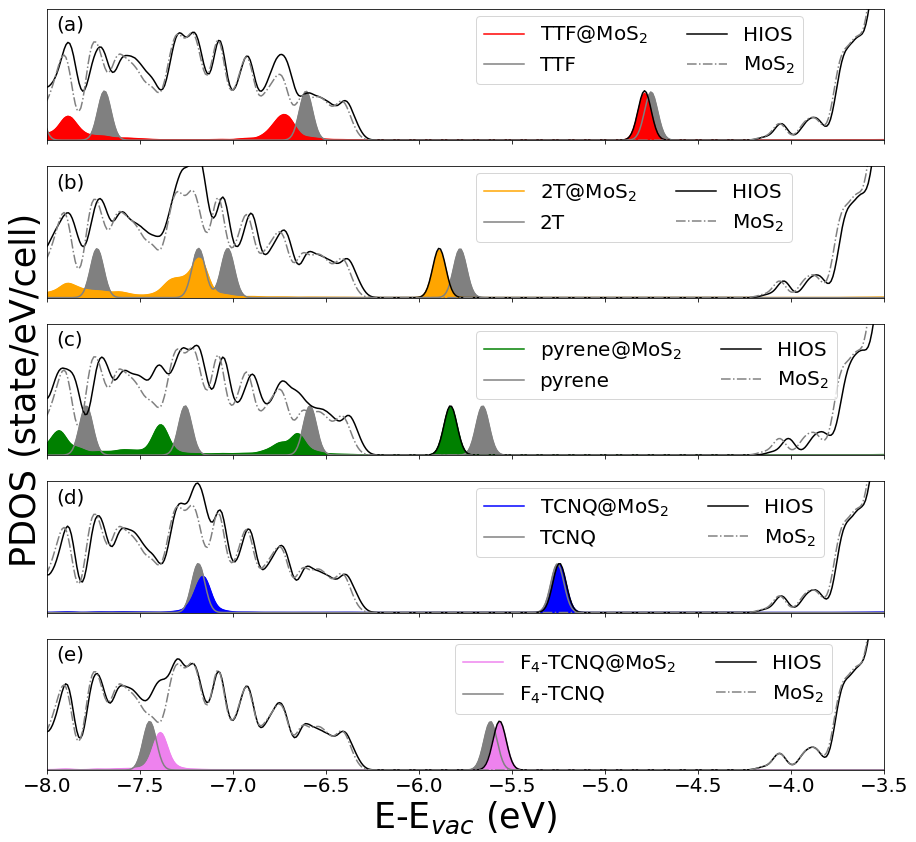}
    \caption{Projected density of states of all \ce{MoS2}-based hybrid inorganic-organic systems (HIOS, black solid lines), including (a)-(b) the donor molecules TTF and bithiophene, (c) the aromatic molecule pyrene, and (d)-(e) the molecular acceptors, TCNQ and \ce{F4-TCNQ}, calculated at the HSE06+D3 level of theory and compared against the results obtained for the isolated constituents shown by dashed lines (\ce{MoS2}) and gray areas (molecules). The contributions of the molecules within the hybrid interfaces are depicted by colored areas. A broadening of 50 meV is applied in all plots. The energy scale is offset to the vacuum level ($E_{vac}$).}
\label{Fig:PDOS-MoS2}
\end{figure}

The PDOS calculated for the \ce{MoS2}-based interfaces and reported in Figure~\ref{Fig:PDOS-MoS2} illustrate well the distribution of the molecular states of the adsorbates with respect to the electronic bands of the substrate. 
In the occupied region, hybridization between \ce{MoS2} states and molecular orbitals can be seen especially for the interfaces including the donor molecules and pyrene (Fig.~\ref{Fig:PDOS-MoS2}a-c).
This effect manifests itself as a broadening of the peaks associated with molecular states, which are no longer $\delta$-like maxima as in the isolated counterpart.
On the other hand, acceptor molecules do not exhibit any signs of hybridization with the \ce{MoS2} bands, at least in the energy window displayed in Fig.~\ref{Fig:PDOS-MoS2}d-e.

Additionally, in Fig.~\ref{Fig:PDOS-MoS2}, the consequences of charge transfer between the molecule and the substrate can be seen, as already discussed for the hBN-based heterostructures.
When the donor molecules TTF and 2T are adsorbed on \ce{MoS2}, their energy levels are downshifted with respect to their counterparts in the isolated moieties.
The size of such shifts is not rigid. 
In TTF@\ce{MoS2}, the highest-occupied state coinciding with the HOMO of the molecule is only 51~meV below the highest-occupied orbital of the gas-phase donor.
On the other hand, the HOMO-1 and the HOMO-2 are downshifted by 130~meV and 200~meV, respectively.
In the PDOS of the 2T@\ce{MoS2} interface, the HOMO is downshifted by 110~meV with respect to the gas-phase counterpart.
For the HOMO-1, HOMO-2, and HOMO-3, the shift of the molecular levels due to charge transfer is entangled with the hybridization with \ce{MoS2} bands, which induces a remarkable smearing in the corresponding peaks.
As a result, a quantitative assessment of the former effect is not straightforward.
A similar behavior is shown also by the PDOS of the pyrene@\ce{MoS2} heterostructure, whereby, interestingly, the downshift of the HOMO is the largest among those seen in Fig.~\ref{Fig:PDOS-MoS2}.
In this system, the three occupied states of pyrene that are visible in Fig.~\ref{Fig:PDOS-MoS2}c are also subject to the joint action of charge-transfer-induced downshift and hybridization with \ce{MoS2} bands.
The energy levels of the molecular acceptors adsorbed on \ce{MoS2} are upshifted by the creation of an interfacial dipole with the monolayer.
Similar to the scenario offered by the hBN-based heterostructures, the magnitude of this effect is much less pronounced than for the donors and signs of hybridization with the substrate bands are hardly visible. 
In the TCNQ@\ce{MoS2} heterostructure (Fig.~\ref{Fig:PDOS-MoS2}d), the molecular levels are essentially aligned with their counterparts in the isolated molecule.
The PDOS of the \ce{F4-TCNQ}@\ce{MoS2} interface exhibits a similar behavior (Fig.~\ref{Fig:PDOS-MoS2}d) but, in this case, the upshift of the HOMO and LUMO levels of \ce{F4-TCNQ} is almost rigid and as large as 60~meV. 

\section{Summary and Conclusions}
In summary, we presented a DFT study of hybrid interfaces formed by hBN and \ce{MoS2} monolyers acting as substrates for five physisorbed molecules: two electron-donor species, TTF and 2T, two acceptors, TCNQ and \ce{F4-TCNQ}, and the aromatic hydrocarbon, pyrene. 
All molecules adsorb substantially flat on both substrates, although structural modifications can be seen depending on the chemical nature of adsorbates and substrates:
Donor and acceptor compounds undergo minor distortions due to the presence of S and N atoms therein, respectively; hBN ripples slightly when interacting with the physisorbed molecules while, owing to its larger rigidity, the structure of \ce{MoS2} in the hybrid interfaces is unchanged compared to the free-standing configuration.
From an energetic point of view, all material combinations form stable heterostructures thanks to the contribution of dispersive interactions, which are quantitatively accounted for in our calculations. 
As a general trend, pyrene and the acceptors adsorb more favorably on both substrates than the considered donors.

From the analysis of the electronic structure, we noticed weak coupling between molecules and hBN, as expected from the chemically inert and insulating character of this 2D material.
In the considered hBN-based heterostructures, both type-I and type-II level alignments are formed. 
Straddling lineups appear for the donor molecules, TTF and 2T, and with pyrene; staggered ones are driven by the acceptors TCNQ and \ce{F4-TCNQ} and their relatively low frontier levels with respect to the vacuum.
In contrast, all \ce{MoS2}-based hybrid systems exhibit a type-II level alignment, with the highest-occupied (lowest-unoccupied) level of the interface coinciding with the HOMO (LUMO) of the electron-donating (-withdrawing) molecule.
The projected density of states of all considered interfaces show two concomitant effects: (i) hybridization between the electronic states of the inorganic and organic components, involving only marginally the frontier orbitals of the physisorbed molecules and (ii) charge transfer between the molecules and the monolayer substrates shifting the molecular energy levels up- or downwards, depending on the electron-donating or electron-withdrawing nature of the organic compounds.
Interestingly, both effects are qualitatively and, to a large extent, also quantitatively similar regardless of the substrate. 

The results to this work provide important indications to rationalize the design of low-dimensional hybrid interfaces for opto-electronic applications.
Our findings suggest that the characteristics of the physisorbed molecules play a bigger role in determining the details of the electronic structure of the heterostructure than those of the inorganic substrate.
However, the band-gap of the latter and the relative energies of their band edges rules to the largest extend the level alignment of the hybrid system.
Future work on the characterization of the electronic excitations is expected to supplement this analysis for a deeper understanding of the opto-electronic activity of these novel materials. 

\section*{Author Contributions}
\textbf{Giacomo Melani:} Investigation, Data Curation, Visualization, Writing - Original Draft; \textbf{Juan Pablo Guerrero:} Investigation, Data Curation, Visualization, Writing - Original Draft; \textbf{Ana M. Valencia:} Data Curation, Visualization, Supervision, Writing - Original Draft; \textbf{Jannis Krumland:} Supervision, Writing - Review \& Editing; \textbf{Caterina Cocchi:} Conceptualization, Supervision, Project administration, Funding acquisition, Writing - Review \& Editing; \textbf{Marcella Iannuzzi:} Conceptualization, Supervision, Project administration, Funding acquisition, Writing - Review \& Editing.

\section*{Data Availability Statement}
All data produced in this work are available free of charge at 10.5281/zenodo.6388531.

\section*{Conflicts of interest}
There are no conflicts to declare.

\section*{Acknowledgements}
This work was carried out in the framework of the strategic partnership between the University of Z\"{u}rich and the Humboldt Universit\"{a}t zu Berlin. 
G.M. and M.I. gratefully acknowledge computational support from the Swiss National Supercomputing Centre (CSCS) under project s965 ``Molecules at interfaces from density functional theory''. G.M. acknowledges funding from the University of Z\"{u}rich Forschungskredit Postdoctoral Fellowship.
J.P.G, A.M.V., J.K., and C.C. appreciate funding from the German Research Foundation (DFG), project number 182087777 -- CRC 951, and computational resources from the North-German Supercomputing
Alliance (HLRN), project bep00104. Additional support is acknowledged by A.M.V. and C.C. to the German Federal Ministry of Education and Research (Professorinnenprogramm III), and by the State of Lower Saxony (Professorinnen für Niedersachsen).


\providecommand{\latin}[1]{#1}
\makeatletter
\providecommand{\doi}
  {\begingroup\let\do\@makeother\dospecials
  \catcode`\{=1 \catcode`\}=2 \doi@aux}
\providecommand{\doi@aux}[1]{\endgroup\texttt{#1}}
\makeatother
\providecommand*\mcitethebibliography{\thebibliography}
\csname @ifundefined\endcsname{endmcitethebibliography}
  {\let\endmcitethebibliography\endthebibliography}{}

\end{document}